\documentclass[aps,prl,reprint,superscriptaddress,longbibliography]{revtex4-2}
\usepackage[T1]{fontenc}
\usepackage{float}
\usepackage{amsmath,amssymb,bm,graphicx}
\usepackage[colorlinks=true,citecolor=blue,linkcolor=blue,urlcolor=blue]{hyperref}
\newcommand{\E}{\mathbb{E}}
\newcommand{\Is}{I_{\mathrm{sense}}}
\newcommand{\Ia}{I_{\mathrm{act}}}
\newcommand{\TE}{\operatorname{TE}}
\newcommand{\SM}[1]{Supplemental Material, Sec.~S#1}
\newcommand{\SIFig}[1]{Fig.~S#1}
\usepackage{etoolbox}
\AtBeginEnvironment{equation}{\ifdefined\linenomath\linenomath\fi}
\AtEndEnvironment{equation}{\ifdefined\endlinenomath\endlinenomath\fi}
\AtBeginEnvironment{align}{\ifdefined\linenomath\linenomath\fi}
\AtEndEnvironment{align}{\ifdefined\endlinenomath\endlinenomath\fi}
\begin{document}
\title{Information-theoretic bounds and costs of antithetic control}
\author{Ga\v{s}per Tka\v{c}ik}
\email{gasper.tkacik@ist.ac.at}
\affiliation{Institute of Science and Technology Austria, AT-3400 Klosterneuburg, Austria}
\affiliation{Tokyo College, University of Tokyo, Bunkyo-ku, 113-0033 Tokyo, Japan}
\author{Praful Gagrani}
\affiliation{Institute of Industrial Science, The University of Tokyo, 4-6-1 Komaba, Meguro-ku, Tokyo 153-8505, Japan}
\author{Mustafa Khammash}
\affiliation{Department of Biosystems Science and Engineering (D-BSSE), ETH Z\"urich, CH-4058 Basel, Switzerland}

\begin{abstract}
Biological feedback must control a fluctuating system with a fluctuating controller. We study antithetic integral feedback, whose performance depends primarily on the speed of relaxation and recovery after external perturbations. We show that, in the linear-noise approximation, this speed is bounded by directed information transmission rates from the system to the controller (sensing) and from the controller to the system (actuation); an optimal controller jointly minimizes both flows at each prescribed speed. At even smaller copy numbers, discrete sensing and rare actuator jumps produce different information scaling. Molecular and flux costs impose distinct constraints; in the slow limit, fixing these costs selects an optimal allocation between controller species. This work showcases deep connections between stochastic control and information flows in chemical reaction networks.
\end{abstract}
\maketitle
\raggedbottom
\setlength{\intextsep}{6pt}

Regulatory processes in biological systems maintain essential functions even though their measurements of internal state and environment, as well as their own feedback actions, are noisy. At the molecular scale, both the core system and its controller---often difficult to disentangle in evolved systems---fluctuate: sensing, comparison, and actuation are chemical reactions with finite molecular populations and reaction turnover~\cite{Tkacik2025}. Limits on noise suppression~\cite{Lestas2010}, causal analyses of feedback fluctuations~\cite{Kobayashi2016}, cellular information measurements~\cite{Cheong2011,Selimkhanov2014}, and links between information, speed, and control in transcription~\cite{Zoller2026} motivate an information-based account of stochastic control. In a recent landmark study~\cite{Das2026}, Das and ten Wolde extended traditional information flow analyses to directional (i.e., transfer entropy) flows, commensurate with the essential role of feedback in the studied chemotactic setting~\cite{Mattingly2021,Reinhardt2023,Nakamura2026}. Here we ask a similar question in stochastic chemical reaction setting: How much information, carried by which events, is needed to stabilize a given concentration, using only feedback internal to the reaction network?

Antithetic integral feedback (AIF) is one of the simplest homeostatic reaction network motifs. It has been realized in synthetic circuits~\cite{Aoki2019,Frei2022} and proposed as an endogenous mechanism~\cite{Scheepers2023}; it comes with an extensive theory of adaptation, design, and noise~\cite{Gupta2014,Briat2018,Olsman2019,OlsmanArchitecture2019,Kell2023,Filo2022}. We connect AIF's performance to directed information flows and metabolic costs using the reaction-based approach of Moor and Zechner~\cite{MoorZechner2023,Moor2026} to analytically compute the Gaussian information rates, as well as path-weight sampling (PWS)~\cite{Reinhardt2023} and its extensions~\cite{Das2025,MLPWS2025} for work beyond the Gaussian approximation.

\begin{figure*}[t]
\setlength{\hsize}{\textwidth}
\includegraphics[width=\textwidth]{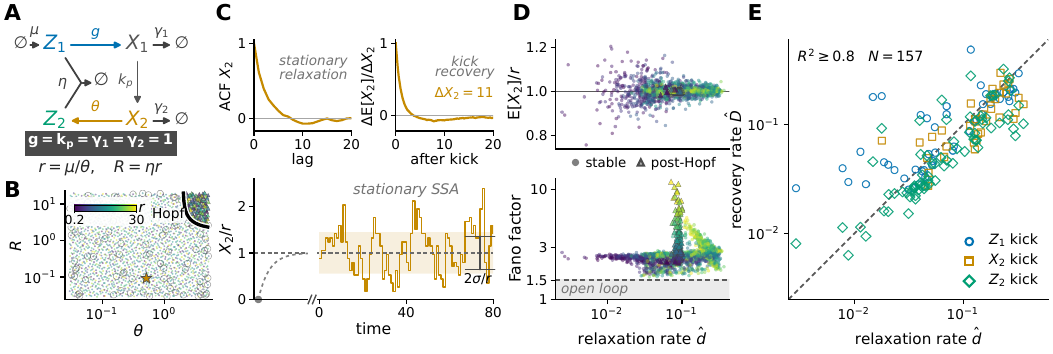}
\caption{\textbf{Controller speed, not stationary variance, is the primary performance measure for stable AIF control.}
\textbf{(A)} Plant $X_1,X_2$: catalytic production $g,k_p$, degradation $\gamma_1,\gamma_2$, all set to one. Controller $Z_1,Z_2$: constants $\mu,\theta,\eta$ combine into $r,R,\theta$ as shown. Time unit: plant lifetime.
\textbf{(B)} Atlas of 2\,048 parameter sets, plotted in the $(\theta,R)$ plane, color: setpoint $r$. Each plotted point is one parameter set. Dots/triangles: stable/post-Hopf deterministic sectors, separated by black curve. Gray circles: 96 parameter sets selected for perturbations; star: example in C.
\textbf{(C)} Characteristics of output species $X_2$. \textit{(Bottom)} Stationary output trace normalized by $r$. Bands show $1\pm\sigma/r$, with $\sigma^2=\operatorname{Var}(X_2)$. Initial gray dashed curve: schematic approach to stationarity. The output autocorrelation timescale \textit{(top left)} is similar to the timescale of recovery after an external, 11-molecule kick to the output \textit{(top right)}.
\textbf{(D)} AIF robustly maintains its setpoint ($\E[X_2]=r$, \textit{top}), with stationary fluctuations quantified by the Fano factor \textit{(bottom)}. Each point: 32 stationary replicate trajectories per parameter set.
\textbf{(E)} Full-state recovery rate $\hat D$ co-varies with the stationary relaxation rate $\hat d$ across 82 AIF parameter sets selected from B perturbed by different kicks (legend; $N=157$ perturbations; $p\simeq10^{-23}$, rank test). Details: \SM{2.3}.
}
\label{fig:phenotype}
\end{figure*}

An AIF controller regulates a two-stage plant $X_1\to X_2$ through actuator $Z_1$ and output-sensing species $Z_2$ [Fig.~\ref{fig:phenotype}A]. Production of $Z_1$ at rate $\mu$, sensing at rate $\theta X_2$, and mutual annihilation at rate $\eta Z_1Z_2$ integrate the production imbalance to maintain the output at a fixed setpoint~\cite{Briat2016}. We choose equal plant production and degradation rate constants and set them to one. Using this time unit, three positive controller parameters remain: the setpoint $r=\mu/\theta$, which also sets the mean population; the sensing rate $\theta$ and annihilation scale $R=\eta r$, which together tune controller kinetics. Exact stationary balances give $\E[Z_1]=\E[X_1]=\E[X_2]=r$; the deterministic fixed point additionally has $Z_2^*=r\theta/R$ (\SM{1}~\cite{Supplemental}).

We first survey AIF behavior beyond its mean setpoint, by generating 2\,048 parameter sets with Sobol sampling across logarithmic $(R,\theta,r)$ parameters, on both sides of the deterministic Hopf boundary [Fig.~\ref{fig:phenotype}B; \SM{2}]. A stationary trajectory, its autocorrelation, and recovery after an external output perturbation illustrate the resulting fluctuations [Fig.~\ref{fig:phenotype}C]. Across the sampled parameter sets, mean adaptation coexists with  output Fano factors above the matched open-loop value $3/2$ [Fig.~\ref{fig:phenotype}D]. Outside the unstable post-Hopf sector (which we do not further analyze in detail), stationary noise varies only modestly with our choice of controller parameters: the central 90\% of Fano factors lie between $1.9$ and $3.4$, less than a twofold range, compared with a thirteenfold range in relaxation rate variability.

Therefore, changing AIF's parameters surprisingly does not seem to exert a major impact on its precision (i.e., variance around the setpoint) -- a natural performance measure for a generic homeostatic controller. Instead, the relaxation rate emerges as the more relevant characteristic. This is because AIF's actuator supplies all plant input while AIF controller reactions generate feedback as well as noise: slowing down these reactions can prolong fluctuations without greatly increasing the variance, in contrast to typical setups where control would weaken against a fixed external noise source (\SM{2.1}).

The intrinsic relaxation rate $d$ describes the slowest stationary decay mode; in the linear-noise approximation (LNA), $d=-\max_i\operatorname{Re}\lambda_i$ for closed-loop poles $\lambda_i$. Its estimate $\hat d$ correlates strongly with recovery rate $\hat D$ after an external kick [Fig.~\ref{fig:phenotype}E; \SIFig{1} and \SM{2} detail rate extraction and the differences between intrinsic relaxation and finite-kick recovery]. We therefore use the intrinsic relaxation rate $d$ as controller speed and our main AIF performance measure.

\begin{figure*}[tb]
\setlength{\hsize}{\textwidth}
\includegraphics[width=\textwidth]{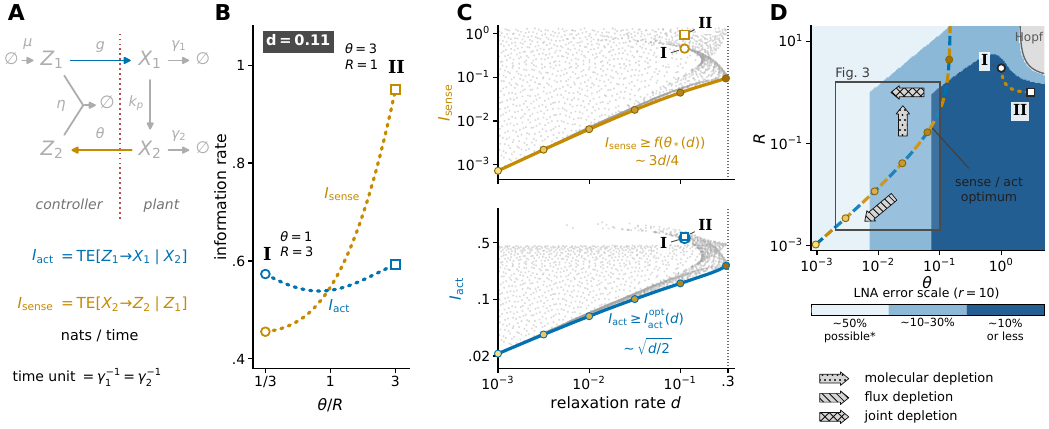}
\caption{\textbf{In the linear-noise approximation, a single controller jointly minimizes sensing and actuation information flow at a given speed.}
\textbf{(A)} Sensing $\Is$ (ochre) and actuation $\Ia$ (blue) are conditional transfer-entropy rates between plant and controller; red dotted line: plant--controller boundary.
\textbf{(B)} A constant-$d$ family links I and II, demonstrating different information flows at equal speed; endpoints share all poles.
\textbf{(C)} Sampling of 2\,565 stable LNA parameter sets (gray dots). Solid curves: minimum sensing \textit{(top)} and actuation \textit{(bottom)} information rates at each required speed. Filled markers: optimal controllers at chosen $d$ (color), matched to the optimal line in D. Dotted vertical: LNA ceiling $d=1/3$.
\textbf{(D)} Alternating blue--ochre line: optimal controllers jointly minimizing $\Ia$ and $\Is$, parametrized by speed $d$ (dot color). Dotted line: I--II path (B). Gray region: deterministic post-Hopf unstable region. Blue bands provide notional LNA error scales at $r=10$ (\SM{4}). Patterned arrows: molecular, flux, and joint depletion routes out of LNA (see text). Rectangle: slow-parameter region examined in Fig.~\ref{fig:discrete}.}
\label{fig:lna}
\end{figure*}

Information enters the controller through sensing and returns through actuation [Fig.~\ref{fig:lna}A]. We measure conditional transfer-entropy rates,
\begin{equation}
 \begin{aligned}
 \Is&=\TE[X_2\to Z_2\mid Z_1],\\
 \Ia&=\TE[Z_1\to X_1\mid X_2],
 \end{aligned}
 \label{eq:cuts}
\end{equation}
in nats per plant lifetime. These rates quantify how much the source history ($X_2$ for $\Is$, $Z_1$ for $\Ia$) improves prediction of the receiving group's reaction events beyond its own history: the full controller history $(Z_1,Z_2)$ for sensing and the full plant history $(X_1,X_2)$ for actuation. The two measures separate both feedback directions, which the ordinary mutual information rate cannot discriminate~\cite{MoorZechner2023,Moor2026}. Dedicated PWS Monte Carlo can estimate information rates for the exact stochastic model~\cite{Reinhardt2023,Das2025} at selected parameter sets (\SM{8}; \SIFig{6}).

A fixed relaxation rate, estimated by $\hat d$ in simulations, can be realized by many parameter combinations with different information flows. For example, in the LNA we can trace a family of controllers at fixed intrinsic rate $d$ for which sensing information changes more than twofold [Fig.~\ref{fig:lna}B]. Even the endpoints I and II, related by $\theta\leftrightarrow R$, differ this much although they share all poles: their linearized decay rates and oscillation frequencies match, as does their whole-loop sensitivity to disturbances. These deterministic properties, however, do not fix stochastic fluctuation amplitudes or information rates (\SM{1}).

Rapid recovery requires prompt, reliable detection of departures from the setpoint and prompt, reliable actuation to correct them. The two information rates quantify transmission along these successive steps, motivating the question: Is a minimum sensing or actuation information rate required to support a given speed? Both rates are analytically computable for every stable parameter set using reaction-based LNA filtering~\cite{MoorZechner2023,Moor2026} (\SM{1}):
\begin{equation}
 \begin{aligned}
 2\Is^{\rm LNA}&=\sqrt{2+2\theta+2\sqrt{1+4\theta}}-2,\\
 2\Ia^{\rm LNA}&=\sqrt{(\theta+R)^2+2\theta+2R\sqrt{2\theta}}-\theta-R.
 \end{aligned}
 \label{eq:lna_rates}
\end{equation}
These Gaussian rates are independent of $r$ at fixed $(\theta,R)$; exact CME rates need not be. We minimize each information rate at required relaxation rate $d$: $\mathcal I_q(d)=\min_{\theta,R>0:\,d(\theta,R)\ge d}I_q$, for $q=\mathrm{sense},\mathrm{act}$. We prove that both minima are attained on the same curve $(\theta_*(d),R_*(d))$, parametrically traced out for $0<d<1/3$ [Fig.~\ref{fig:lna}C]. At each point on this optimal curve, two slow poles coincide at $-d$; spectral constraints and monotonicity of the information rates establish global optimality within the stable LNA (\SM{3}).

Along this optimal curve, $\theta_*\sim R_*\sim d$ as the required speed $d\to0$, but the minimum information rates have different asymptotes:
\begin{equation}
 \begin{gathered}
 I_q\ge I_q(\theta_*(d),R_*(d))=\mathcal I_q(d),\\
 \mathcal I_{\mathrm{sense}}(d)\sim\frac34d,
 \qquad \mathcal I_{\mathrm{act}}(d)\sim\sqrt{\frac d2}.
 \end{gathered}
 \label{eq:lna}
\end{equation}
The minimum sensing rate is thus parametrically smaller than the minimum actuation rate. Taken together, our first key result is the existence of an \emph{optimal LNA bound}, on which a single controller jointly minimizes both information rates at each prescribed speed.

\begin{figure*}[tb]
\setlength{\hsize}{\textwidth}
\includegraphics[width=\textwidth]{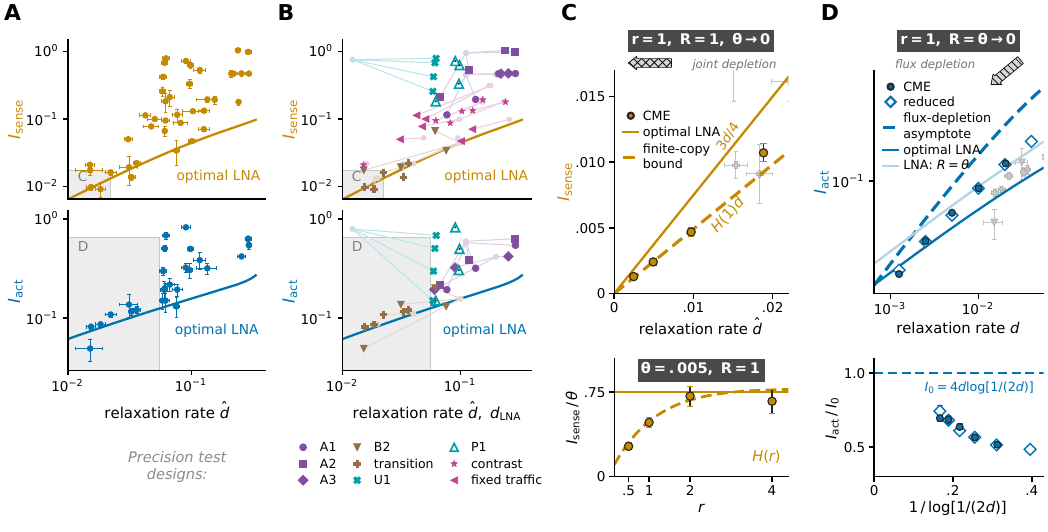}
\caption{\textbf{Single-molecule and rare-reaction limits change the AIF information scaling.}
\textbf{(A)} 40 sensing \textit{(top)} and 26 actuation \textit{(bottom)} TE-PWS estimates. Vertical error bars: standard error (SE) over 6--48 independent Monte Carlo trajectories. Horizontal: bootstrap SE for relaxation rate. Solid curves: optimal LNA bounds. Gray regions: expanded in C (sensing) and D (actuation).
\textbf{(B)} Precision test designs (color/shape, legend below) used in A systematically test deviations of LNA predictions (pale colors) from exact PWS information-rate estimates (strong colors, linked). Eight open upward triangles are post-Hopf, without LNA endpoints.
\textbf{(C)} \textit{(Top)} Joint depletion ($\theta\to0$) changes sensing information. Reduced bound $H(1)d$ (dashed) versus LNA $3d/4$ (solid). \textit{(Bottom)} $\Is/\theta$ compared with discrete $H(r)$ (dashed) and LNA $3/4$ (solid) as $r$ decreases.
\textbf{(D)} \textit{(Top)} Flux depletion ($R=\theta\to0$) changes actuation information. Circles: full-CME estimates. Diamonds: reduced-system estimates (\SM{6}). Dark/light blue lines: optimized/fixed-path LNA. Dashed line: $I_0=4d\log[1/(2d)]$. \textit{(Bottom)} Approach to $\Ia/I_0=1$ as speed $d$ decreases.
Vertical error bars for C, D: SE across $\geq 8$ independent replicates. Horizontal: finite-state approximation error. Details: \SM{8}.}
\label{fig:discrete}
\end{figure*}

At fixed finite $r$, approaching this slow optimum eventually invalidates the Gaussian approximation [Fig.~\ref{fig:lna}D]. We devised an approximate warning score for the breakdown of LNA validity by heuristically combining molecular counts, fluctuations, and controller events per plant lifetime, and calibrated its broad error scales against exact PWS (\SM{4}; \SIFig{2}). Three arrows illustrate different exits from LNA: increasing $R$ at fixed $\theta$ depletes the mean-field sensor population $r\theta/R$ at fixed flux (upward dotted arrow); decreasing $R$ and $\theta$ together at fixed ratio preserves the mean-field populations while making controller reactions rare (down-left striped arrow); decreasing $\theta$ at fixed $R$ causes both forms of depletion (leftward checkered arrow).

Precision tests comparing LNA and exact PWS results [Fig.~\ref{fig:discrete}A,B] guide a closer analysis of the slow-parameter region marked by the rectangle in Fig.~\ref{fig:lna}D. Most sampled controllers lie above the optimal LNA bounds [Fig.~\ref{fig:discrete}A], but their information rates lie predominantly below their own LNA predictions [Fig.~\ref{fig:discrete}B]. For example, at $(\theta,R,r)=(1,3,.25)$, sensing information is about $75\%$ lower, while the empirical relaxation rate is only $31\%$ lower. Finite-copy effects therefore change information and speed by different amounts. We analyze two slow-controller limits to identify the underlying mechanisms.

For sensing, consider joint depletion: $\theta\to0$ at fixed finite $r$ and fixed positive $R=O(1)$ [Fig.~\ref{fig:discrete}C]. Sensor molecules become scarce and controller reactions become rare. The actuator remains at a fixed integer population while the plant equilibrates. Averaging the sensing information rate over these long intervals gives
\begin{equation}
 \begin{aligned}
 \Is&=\theta H(r)+o(\theta),\\
 H(r)&=\E_{n\sim\mathrm{Pois}(r)}\E_{\pi_n}
       [X_2\log(X_2/n)],
 \end{aligned}
 \label{eq:sense}
\end{equation}
where $\pi_n$ is the stationary plant distribution with the actuator held at $Z_1=n$ (\SM{5}). The contribution at $n=0$ is zero because the plant is then empty (\SIFig{3}). At $r=1$, PWS estimates approach the finite-copy prediction $H(1)d$ as sensing slows [Fig.~\ref{fig:discrete}C, \textit{top}]; its coefficient $H(1)\simeq0.491$ is about $35\%$ below the LNA value $3/4$.
\begin{figure*}[t]
\setlength{\hsize}{\textwidth}
\includegraphics[width=\textwidth]{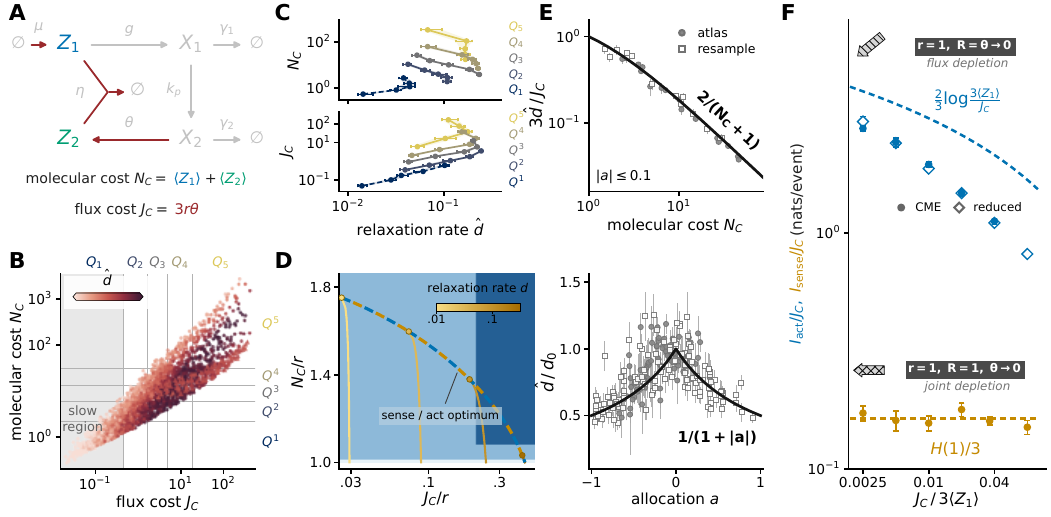}
\caption{\textbf{Molecular and flux costs constrain control performance.}
\textbf{(A)} Molecular cost $N_C$: mean controller population. Flux cost $J_C$: event rates in red channels.
\textbf{(B)} Atlas cost plane shows a sampling over 2\,048 parameter sets. Color: speed $\hat d$. Symbols as in Fig.~\ref{fig:phenotype}B. Gray lines: cost quantiles.
\textbf{(C)} Molecular cost \textit{(top)} or flux cost \textit{(bottom)} versus median speed, within quantiles of the other cost. Error bars and shading: 95\% bootstrap intervals for median speed. Lowest quantile additionally dashed.
\textbf{(D)} LNA cost tradeoffs at fixed $r,d$. Alternating blue--ochre line: common information optimum and minimum-flux endpoint [Fig.~\ref{fig:lna}D]. Along each Pareto branch, reducing molecular cost requires increasing flux cost. Blue LNA validity regions as in Fig.~\ref{fig:lna}D.
\textbf{(E)} Reduced allocation laws (black) compared with stochastic simulation measurements: 79 atlas sets with $\theta,R\le.1$, measured $N_C\ge1$ (circles), and 127 prospective resamples (squares), selected without information optimization (\SM{7}). \textit{(Top)} Nearly balanced subset, $|a|\le.1$ (15 atlas, 14 resampled). \textit{(Bottom)} All 206 sets. Here $a=(N_1-N_2-1)/(N_C+1)$ and $d_0=2J_C/[3(N_C+1)]$; error bars: 1 SE.
\textbf{(F)} Information per controller event at $r=1$. Circles: full CME using PWS. Diamonds: reduced CME. Error bars: 1 SE. Dashed lines: slow-limit scaling predictions. Hatched arrows as in Fig.~\ref{fig:lna}D. Details: \SM{7} and S8.}
\label{fig:cost}
\end{figure*}
A separate population sweep at $\theta=0.005$, $R=1$ shows $\Is/\theta$ falling as $r$ decreases, following the discrete coefficient $H(r)$ rather than the constant LNA prediction [Fig.~\ref{fig:discrete}C, \textit{bottom}]. Even arbitrarily slow reactions retain the effects of integer populations and empty-actuator intervals.

For actuation, we follow flux depletion: $R=\theta\to0$ at fixed finite $r$ [Fig.~\ref{fig:discrete}D]. Both controller rates vanish together, preserving finite populations while making jumps rare. A broader rare-jump argument suggests the full chemical master equation (CME) information scaling as $\theta\to0$,
\begin{equation}
 \Ia=2r\theta\log(1/\theta)
       +o[\theta\log(1/\theta)].
 \label{eq:act}
\end{equation}
The step-by-step proof outline involves sophisticated stochastic process formalism~(\SM{6}) and awaits expert review. Physical intuition considers actuator births and deaths each occurring at mean rate $r\theta$; between jumps, the plant has time to distinguish successive actuator populations, locating these rare jumps within its long waiting intervals at $\log(1/\theta)$ nats per jump to leading order, yielding the $\Ia$ estimate.

On the plotted $r=1$, $R=\theta$ path in Fig.~\ref{fig:discrete}D, the reduced process has $d=\theta/2$ (\SIFig{4}), giving $I_0=4d\log[1/(2d)]$ (dashed line). The dark-blue optimized LNA bound instead scales as $\sqrt{d/2}$; the light-blue LNA prediction on the same $R=\theta$ path scales as $\sqrt d$. Full-CME and reduced PWS estimates agree. Their ratios $\Ia/I_0$ rise toward unity as speed decreases [Fig.~\ref{fig:discrete}D, \textit{bottom}], but remain appreciably below one: convergence to the leading logarithmic scaling is slow.
Taken together, our second key finding is that single-molecule and rare-reaction regimes beyond LNA produce qualitatively different information scalings, reflecting the importance of individual events for controller operation.

What metabolic resources support these information flows and performance? Molecules store controller state; reactions update it. Their mean population and event rate define molecular and flux costs [Fig.~\ref{fig:cost}A]:
\begin{equation}
 N_C=\E[Z_1+Z_2],\qquad J_C=3r\theta.
 \label{eq:cost}
\end{equation}
The flux identity is exact: reference production, sensing, and annihilation each carry stationary flux $r\theta$; neither cost is entropy production (\SM{7}).

To gain insight, we first examine the AIF controller atlas on a broad cost plane [Fig.~\ref{fig:cost}B]. Within molecular-cost quantiles, speed typically rises with flux, but this relationship can surprisingly invert at high flux. Within flux quantiles, larger molecular populations often slow relaxation [Fig.~\ref{fig:cost}C]. These broadband comparisons, however, mix setpoints and kinetics; to isolate cost tradeoffs, we next fix $r$ and $d$.

Our third key finding is a Pareto front in molecular and flux costs for LNA controllers at fixed setpoint $r$ and speed $d$, with the common information optimum at its minimum-flux endpoint [Fig.~\ref{fig:cost}D; \SM{7}]. Moving toward sensor depletion lowers molecular cost but raises flux cost and both information rates. At small $d$, molecular cost falls from approximately $2r$ at the optimum toward $r$, although the vanishing sensor population eventually invalidates LNA at finite $r$. Controllers with larger sensor populations than the optimum cannot lower flux cost and are dominated. (Other objectives, such as stationary precision, would shift this balance; \SM{7.5}, \SIFig{5}.)

How should a fixed population of $N_C$ molecules be divided between actuator and sensor? For $\theta\to0$, $R=\alpha\theta$ at fixed $\alpha>0$, averaging over the plant gives the exact reduced-CME allocation law [Fig.~\ref{fig:cost}E]
\begin{equation}
 d_{\mathrm{red}}=\frac{J_C}{3\max\{N_1,N_2+1\}},
 \qquad N_i=\E[Z_i].
 \label{eq:allocation}
\end{equation}
At fixed $N_C\ge1$ and $J_C$, the fastest allocation equalizes the denominator's two terms: $N_1=N_2+1$, yielding $d_{\max}=2J_C/[3(N_C+1)]$. This optimization permits the setpoint $r=N_1$ to vary. The decrease of $d_{\max}$ with $N_C$ at fixed $J_C$ makes the cost of slower turnover explicit: even optimally allocated extra molecules reduce the attainable speed. This provides a possible explanation for atlas trends reported in Fig.~\ref{fig:cost}C. The reduced law predicts a universal imbalance curve, $d_{\mathrm{red}}/d_{\max}=1/(1+|a|)$, showing how unequal allocation further slows control [Fig.~\ref{fig:cost}E, \textit{bottom}; \SM{7}].

Finally, the slow-limit scaling predicts different information per controller event [Fig.~\ref{fig:cost}F]: at fixed $r$, $H(r)/(3r)$ for sensing, but growth as $(2/3)\log(3r/J_C)$ for actuation. An isolated actuator jump, unlike an isolated sensing event, leaves a sustained change in plant production that localizes the jump time within an increasingly long waiting interval (\SM{5}--S7).

AIF offers an elementary model of molecular control, analogous in role to the harmonic oscillator in physics. In LNA, one controller minimizes both information rates at each speed and selects the minimum-flux endpoint of the molecular--flux cost tradeoff. Beyond LNA, discrete populations and rare events change information scaling and information per controller event. A global full-CME information--speed bound and extensions to other architectures~\cite{Filo2026} remain open (\SM{9}), alongside sharper LNA-validity criteria and information-rate estimates beyond LNA without full stochastic simulation, potentially using large-deviation theory. Our results also suggest that synthetic AIF~\cite{Aoki2019,Frei2022} may be able to exploit sparse, few-molecule operation for low-cost control while meeting useful speed and precision targets.

\textit{Data availability.} Code to reproduce all computations, PWS sampling, plots, as well as Markdown derivations, are available at GitLab~\cite{code}.

\begin{acknowledgments}
\emph{Acknowledgments.} We thank Pieter Rein ten Wolde and Tetsuya J Kobayashi for constructive comments on the manuscript. GT acknowledges the hospitality of the Kobayashi group during sabbatical stay at the University of Tokyo. This work is partially supported by JSPS (25H01365) and JST CREST (JPMJCR25Q2).
AI assistance through Codex contributed substantively to proofs, computation, and manuscript preparation (\SM{10}).
\end{acknowledgments}

\bibliography{references}
\end{document}